\documentstyle[amssymb,epsf,cite,11pt]{article}

\newcommand{\be}{\begin{equation}}
\newcommand{\ee}{\end{equation}}

\begin{document}

\title{\bf Topological properties and fractal analysis of recurrence network constructed from fractional Brownian motions}

\author{ Jin-Long Liu$^{1}$, Zu-Guo Yu$^{1,2}$\thanks{
  Corresponding author, email: yuzg1970@yahoo.com}  ~and Vo Anh$^{2}$\\
{\small$^1$ Hunan Key Laboratory for Computation and Simulation in
Science and Engineering and }\\
{\small Key Laboratory of Intelligent Computing and Information
Processing of Ministry of Education,}\\
{\small Xiangtan University, Xiangtan,  Hunan 411105, China.}\\
{\small $^{2}$School of Mathematical Sciences, Queensland University of Technology,}\\
{\small GPO Box 2434, Brisbane, Q4001, Australia.}
 }
\date{}
\maketitle

\begin{abstract}
Many studies have shown that we can gain additional information on
time series by investigating their accompanying complex networks.
In this work, we investigate the fundamental topological and
fractal properties of recurrence networks constructed from
fractional Brownian motions (FBMs). First, our results indicate
that the constructed recurrence networks have exponential degree
distributions; the average degree exponent $<\lambda>$ increases
first and then decreases with the increase of Hurst index $H$ of
the associated FBMs; the relationship between $H$ and $<\lambda>$
can be represented by a cubic polynomial function. We next focus
on the motif rank distribution of recurrence networks, so that we
can better understand networks at the local structure level. We
find the interesting superfamily phenomenon, i.e. the recurrence
networks with the same motif rank pattern being grouped into two
superfamilies. Last, we numerically analyze the fractal and
multifractal properties of recurrence networks. We find that the
average fractal dimension $<d_{B}>$ of recurrence networks
decreases with the Hurst index $H$ of the associated FBMs, and
their dependence approximately satisfies the linear formula
$<d_{B}> \approx 2 - H$, which means that the fractal dimension of
the associated recurrence network is close to that of the graph of
the FBM. Moreover, our numerical results of multifractal analysis
show that the multifractality exists in these recurrence networks,
and the multifractality of these networks becomes stronger at
first and then weaker when the Hurst index of the associated time
series becomes larger from 0.4 to 0.95. In particular, the
recurrence network with the Hurst index $H=0.5$ possess the
strongest multifractality. In addition, the dependence
relationships of the average information dimension $<D(1)>$ and
the average correlation dimension $<D(2)>$ on the Hurst index $H$
can also be fitted well with linear functions. Our results
strongly suggest that the recurrence network inherits the basic
characteristic and the fractal nature of the associated FBM
series.
\end{abstract}



\section{Introduction}

Methods of nonlinear time series analysis have been widely applied
in physics, physiology, finance and biology. Complex network
theory has become one of the most important developments in
statistical physics \cite{Albert02}. Recent studies have shown
that complex network theory may be an effective method to extract
the information embedded in time series \cite{Donner12, Donner10}.
Many complicated dynamics systems in nature and society can be
described by complex networks. In a complex system, its elements
are represented by nodes, and their interactions are represented
by directed or undirected edges.

Based on the small-world and scale-free properties of complex
networks \cite{Watts98, Barabasi99},  Newman \textit{et al.}
\cite{Newman03} investigated extensively the  structure and
function of real-world complex networks in different areas. The
advancement of network theory provides us with a new perspective
to perform time series analysis \cite{Donner12, Donner10}.
Therefore, we can further understand the structural features and
dynamics mechanism of complex systems by studying the basic
topological properties of networks. Many algorithms have been
proposed to construct different complex networks from time series
\cite{Xu09}, such as visibility graphs \cite{Lacasa08, Luque09},
space state networks \cite{Li08}, recurrence networks
\cite{Donner12, Donner10, Marwan09}, and nearest-neighbor networks
\cite{Xu08, Liu10}.

Recurrence is a basic characteristic of many complex dynamical
systems, which can be used to describe the dynamics behavior of
systems. Eckmann \textit{et al.} \cite{Eckmann87} developed the
method of recurrence plots (RPs) to visualize the recurrence
property of complex dynamical systems. A remarkable advantage of
this method is that it is very suitable for short and even
non-stationary time series. Applications of the method of RPs can
be found in various fields of research such as physiology,
neuroscience, earth sciences, engineering, biology and finances.
In order to quantify the structures of the recurrence plots,
Zbilut and Webber developed the recurrence quantification analysis
(RQA) based on linear and diagonal line structures of recurrence
plots \cite{Zbilut92, Webber94}. Most of these measures of
complexity are available on TOCSY software platform \cite{ht21}.
Cross recurrence plot (CRP) \cite{Zbilut98, Marwan02} and joint
recurrence plot (JRP) \cite{Romano04} are bivariate and
multivariate extensions of the RPs, respectively.

Recently,  Donner \textit{et al.} \cite{Donner12, Donner10} and
Marwan \textit{et al.} \cite{Marwan09} proposed the recurrence
network based on the method of RPs. This novel idea can be
considered as a complementary view on the RQA, allowing us to gain
additional information from corresponding network-theoretic
measures which cannot be obtained by RQA. They considered the
phase space state vectors $X(i)$ (defined in Section II) as nodes
of a complex network and identify the recurrence matrix of the
original time series with the adjacency matrix of an associated
complex network. In this way, an undirected and unweighted network
is represented by a binary adjacency matrix $A_{i,j}$ (defined in
Section II). Recurrence networks have been used to many model
systems such as logistic map, H\'{e}non map, Lorenz system,
R\"{o}ssler system and Bernoulli map \cite{Donner10, Marwan09,
Donges12}, and real economic series \cite{Donner10a}.

It is well known that fractional Brownian motion (FBM) described
by the Hurst index $H$ $(0 < H < 1)$ is a self-similar process
with stationary increments \cite{Mandelbrot68}. The FBM with
$H=1/2$ is in fact the classical Brownian motion. The correlation
between
 increments of the FBM is negative if $0 < H < 1/2$
and positive if $1/2 < H <1$.  The sample paths of FBM series are
relative more rough and variable for small Hurst index $H$, while
the sample paths of FBM series are more smooth for large Hurst
index $H$.

In this work, we try to reveal the relationship between FBM and
its related recurrence network from the perspective of network
structure. This prompted us to further study the fundamental
topological and fractal properties of recurrence networks
constructed from FBMs. We think that applying recurrence network
analysis (RNA) to FBM can yield results which are theoretically
meaningful. The reasons are: (a) Visibility graphs and recurrence
networks are two main classes of methods to map time series to
networks. Recently some works have been done to study the
relationship between the exponents of newly developed
methods(including visibility graphs) and the Hurst index of the
associated FBMs \cite{Lacasa09, Xie11, ZLY2013}. (b) Donges {\it
et al.} \cite{Donges12} have proposed an analytical framework for
RNA of time series of chaotic maps and stochastic processes (such
as uniformly distributed noise and Gaussian noise). FBMs are
generalization of Gaussian noise. (c) Donner {\it et al.}
\cite{Donner10a} applied RNA to study the real economic time
series. And it is well known that FBM has been used as a
theoretical framework to study economic time series
\cite{Mandelbrot68}. In 1999, Riley {\it et al.} \cite{Riley1999}
applied the RQA technique for analyzing center of pressure (COP)
signals and the nonstationarity of the COP is expected from a
consideration of COP trajectories as FBM.

The remaining of this paper is organized as follows. In Section 2,
we adopt the recurrence plot method to construct recurrence
networks. In Section 3, we investigate the basic topological
characteristics of the recurrence networks constructed from FBMs.
We next introduce the random sequential box-covering method
\cite{Kim07} to calculate the fractal dimension of the recurrence
networks in Section 4. In Section 5, we introduce an improved
algorithm newly developed by our group \cite{Libg2014}, which is
based on the modified fixed-size box-counting algorithm
\cite{Wang12}, to probe the multifractal behavior of recurrence
networks. We finally draw some conclusions in Section 6.

\section{Recurrence network}

The method of recurrence plots is based on the theory of phase
space reconstruction introduced by Packard \textit{et al.}
\cite{Packard80}. In order to easily construct a more suitable
space and then to reveal more meaningful information from original
time series, Packard \textit{et al.} \cite{Packard80} proposed the
derivative reconstruction method and time delay method to
reconstruct a finite-dimensional phase space.  However, in
practice, we do not know any prior information of time series,
therefore the latter method is widely used to reconstruct phase
space of time series. For a given scalar time series
$\{x_{i},i=1,2,\cdots,L\}$, we construct the delay vectors
$$ X(i) = (x_{i},x_{i+\tau},\cdots,x_{i+(m-1)\tau}),   X(i)\in {\bf R}^m,  \eqno{(1)} $$
for $i=1,2,\cdots,N$, where $N=L-(m-1)\tau$, $m$ is the embedding
dimension and $\tau$ is the time delay. Takens' embedding theorem
ensures that we can recreate a topologically equivalent
$m$-dimensional phase space from an infinite noise-free time
series data by means of the time delay method \cite{Takens81}. The
most significant advantage of phase space reconstruction is that
it can preserve the geometrical invariants of the original system,
such as the fractal dimension and the Lyapunov exponents
\cite{Takens81}. So we can investigate much about the dynamics
characteristics of time series in phase space more easily. The
basic embedding theorem of phase space reconstruction assumes that
we can choose the delay time $\tau$ without any limitation and the
embedding dimension $m$ under the condition of $m \geq 2d+1$,
where $d$ is the fractal dimension of the underlying attractor.
However, since the observed time series data sets are finite and
noisy in the most common case, the selection of the time delay
$\tau$ and the embedding dimension $m$ is rather important for the
quality of the phase space reconstruction.

The binary recurrence matrix $R_{N N}=(R_{i,j})_{N N}$ is defined
as
$$ R_{i,j} = \Theta(\varepsilon - \|X(i)-X(j)\|),   i,j = 1,2,\cdots,N,  \eqno{(2)} $$
where $\varepsilon$ is a threshold distance, $\|\cdot\|$ is a
suitable norm (e.g., the $L_{1}$ - norm, the Euclidean norm and
the $L_{\infty}$ - norm) in the considered phase space and
$\Theta(x)$ is the Heaviside step function (i.e. $\Theta(x)=0$ if
$x<0$, and $\Theta(x)=1$ otherwise). The resulting matrix $R_{N
 N}$ is a symmetric matrix where the elements $R_{i,j} = 0$
or $1$. Moreover, the matrix $R_{N N}$ exhibits the line of
identity (the main diagonal) $R_{i,i} = 1$. In a two-dimensional
space, the recurrence plot is described by using different colors
for different values of the recurrence matrix, e.g., plotting a
black dot at the coordinates $(i,j)$ if $R_{i,j}=1$, and a white
dot if $R_{i,j}=0$. Thus the recurrence plot always has a black
main diagonal line.

Now we obtain the adjacency matrix $A_{N N}=(A_{i,j})_{N N}$ of
the recurrence network from the recurrence matrix
 $R_{N N}$ as
$$A_{i,j} = R_{i,j}-\delta_{i,j},   \eqno{(3)}$$
where $\delta_{i,j}$ is the Kronecker delta function introduced
here in order to avoid self-loops \cite{Donner10}. In complex
network, each state vector $X(i)$ of the reconstructed phase space
represents a single node. The topological structure of the
recurrence network can be described with the adjacency matrix
$A_{N N}$, and the elements $A_{i,j}=1$ and $A_{i,j}=0$ correspond
to connection and disconnection, respectively. And the size of the
network is $N$.

As mentioned above, the choice of the time delay $\tau$ and the
embedding dimension $m$ plays an important role in the recurrence
plot method. If the time delay $\tau$ selected is too small, each
coordinate of the vector $X(i)$ will be so close to each other
that the trajectories of the reconstructed phase space are
compressed along the identity line, and this phenomenon is
referred to as redundance. If the time delay $\tau$  selected is
too large, the coordinates of the vector $X(i)$ are completely
independent of each other and then the reconstructed attractor
dynamics become causally disconnected, and this phenomenon is
referred to as irrelevance \cite{Casdagli91}. At present,
auto-correlation function \cite{Rosenstein94} and mutual
information method \cite{Fraser86} are the most common approaches
for the estimation of the time delay $\tau$. Fraser \textit{et
al.} \cite{Fraser86} proposed a recursive method of calculating
mutual information and pointed out that the first local minimum of
mutual information is the best criterion for choosing time delay
$\tau$ in phase space reconstruction from time series data.

Takens' embedding theorem gives us a way to estimate the embedding
dimension $m$. This is to say that all self-intersections of the
orbit (which is the reconstructed attractor) in the reconstructed
phase space will be eliminated for any embedding dimension $m \geq
2d+1$. The attractor will be completely unfolded in this embedding
dimension $m$. But in practice, since the dimension $d$ is unknown
in many cases, we cannot directly calculate the embedding
dimension $m$ from the Takens' embedding theorem. Many algorithms
based on computing some invariants on the attractor have been
proposed to calculate the minimum embedding dimension $m$. The
most popular method for calculating the minimum embedding
dimension is the False nearest-neighbors (FNN) algorithm
\cite{Kennel92, Cao97}.  According to the Takens' embedding
theorem, we assume that $m$ is qualified as an embedding
dimension. If any two points are close to each other in the
$m$-dimensional phase space, then they are still close to each
other in the $(m+1)$-dimensional phase space. We call this two
points true neighbors, otherwise, they are false neighboring
points. The main idea of the FNN algorithm is to measure the
percentage of false nearest neighbors along a signal trajectory
change with increasing embedding dimension. The optimal embedding
dimension means that there is no false neighbors and
self-intersections in the phase space. This indicates that the
minimum embedding dimension is that for which the percentage of
false nearest neighbors drops to zero or nearly zero for a given
tolerance level for the first time \cite{Kennel92}.

The threshold distance $\varepsilon$ is another key parameter of
an RP. The selection of the threshold $\varepsilon$ depends
strongly on the system considered. If $\varepsilon$ is too small,
there may be almost no recurrence points and we cannot learn
anything about the recurrence structure of the underlying system.
On the other hand, if $\varepsilon$ is too large, almost every
point is a neighbor of every other point, which leads to a lot of
artifacts and redundant information \cite{Marwan07}. Therefore,
special attention has been required on its choice. In practice,
the most common method is to set  $\varepsilon$ for a certain
proportion of the standard deviation $\sigma$ of the original time
series. However, Riley and Van Orden suggested choosing
$\varepsilon$ such that the recurrence point density remains low
(often smaller than 5\%) \cite{Riley05}. This method is often used
in RQA. The advantage of choosing an adaptive recurrence threshold
by using a fixed recurrence rate in RQA is that it can preserve
the recurrence point density and then allows us to compare the
results of recurrence quantification analysis of different
systems. It is easy to imagine that there is no direct
relationship between recurrence rate or link density and network
connectivity. And it is so difficult for us to find a fixed
recurrence rate that all recurrence networks obtained at
recurrence threshold $\varepsilon$ calculated from the fixed
recurrence rate are connected. However, we restrict our attention
to the consideration of connected network in this article.
Therefor we do not adopt the fixed recurrence rate for obtaining
such an appropriate recurrence threshold. If we do so, then we
will not be able to study the topological and fractal properties
of the entire network. In other word, we can only focus on the
disconnected components or the largest connected component of each
recurrence network.

The critical phenomenon is one of the most interesting findings of
complex networks. In 1959, Erd\H{o}s \textit{et al.}
\cite{Erdos59} introduced the classical random graph model and
studied the structural phase transition of the birth of the giant
connected component in the network architectures. Every pair of
nodes of random graph is connected with probability $p$. They
found that there exists a critical probability $p_{c}$ where a
giant connected component was formed and it can be spread across
the entire network. If the probability $p$ is less than this
critical probability $p_{c}$, the network is composed of
disconnected components. Similarly to the critical behavior of
random-graph theory, we study the critical connectivity of the
recurrence networks by examining the size of the largest connected
component and then determine the critical recurrence threshold
$\varepsilon_{c}$. Here we firstly employ the connection rate of
network to quantify the connectivity of the recurrence network,
which is defined as the number of nodes of the largest connected
component divided by the size of the network. Secondly, we observe
the change of connection rate of network with the increase of the
parameter $\varepsilon$. It can be easily imagined that the
connection rate of network increases when the parameter
$\varepsilon$ becomes larger, as illustrated in Fig. 1. In Fig. 1,
we set the recurrence threshold $\varepsilon$ ranging from
$0.01\sigma$ to $0.3\sigma$ with a step of $\sigma_s=0.01\sigma$,
where $\sigma$ is the standard deviation of the original FBM
series. Finally, we choose an appropriate recurrence threshold
$\varepsilon_{c}$ where the connection rate of the network arrives
at $1$ for the first time. As can be seen in Fig. 1, we set the
critical recurrence threshold $\varepsilon_{c} = 0.12\sigma$. The
recurrence network obtained at recurrence threshold
$\varepsilon_{c}$ will be connected. The threshold $\varepsilon$
beyond this critical value $\varepsilon_{c}$ will result in
redundant connections among nodes, and threshold $\varepsilon$
below this critical value $\varepsilon_{c}$ will lead to a number
of disconnected components where each component only contains a
connected subnetwork. Consequently, we select this suitable
recurrence threshold $\varepsilon_{c}$ to construct the recurrence
network and then investigate its topological and fractal
properties. In addition, we also notice that the critical
recurrence threshold $\varepsilon_{c}$ decreases with the increase
of the Hurst index $H$ from 0.4 to 0.95. We know that the FBM
series is relatively more smooth and regular for large Hurst index
$H$. The consequence is that there are many data points in the
original FBM series are mapped to state vectors which are close to
each other in a dense region of the phase space so that the
largest connected component of the recurrence network contains
more nodes than the one of the network with the small Hurst index
$H$. The specific explanation for this phenomenon can be found in
the second subsection of the Section 3.

In this work, we consider FBM of length $L = 2^{12}$ with
different Hurst indices $H$ ranging from 0.05 to 0.95 (the step
difference is 0.05). Recurrence networks can be constructed by the
following steps:

\begin{enumerate}
\item[(i)] For a given FBM time series, we use the mutual
information method \cite{Fraser86, Roulston99} and the FNN
algorithm \cite{Kennel92} to calculate the time delay $\tau$ and
the embedding dimension $m$, respectively. These two procedures
are available on TOCSY software platform \cite{ht21}. The values
of embedding dimensions $m$ we got vary from 5 to 7, and those of
the time delay $\tau$ vary from 10 to 20.

\item[(ii)] We use the connection rate of the recurrence network
to determine the suitable threshold distance $\varepsilon_{c}$
\cite{Tang12}. We observe the connection rate of network reaching
$1$ for the first time when the parameter $\varepsilon$ ranging
from $0.01\sigma$ to $0.8\sigma$ with a step of
$\sigma_s=0.01\sigma$, where $\sigma$ is the standard deviation of
the original FBM series. The connection rate of network can be
calculated by Matlab-BGL toolbox \cite{Gleich} in Matlab.

\item[(iii)] Based on the above steps, we can calculate the
recurrence matrix $R_{N N}$ by Eq. (2). Then we obtain the binary
adjacency matrix $A_{N N}$ of an unweighted and undirected
recurrence network by Eq. (3). As a result, the recurrence network
is also connected.
\end{enumerate}

For clarity of visualization, we here only draw the recurrence
network for the FBM series of length $L=2^{8}$ with Hurst index
$H=0.6$.  After obtaining the adjacency matrix $A_{N N}$ of
recurrence network according to the above three steps, we convert
the adjacency matrix $A_{N N}$ into ``Pajek" \cite{ht40} format
and then we adopt the ``Pajek" which is a freely available
software platform for visualizing and analyzing complex networks
to visualize the recurrence network. As can be clearly seen in
Fig. 2, the recurrence network constructed from the associated FBM
series is connected.

In addition, we find that the connection rates of recurrence
networks constructed from FBMs with Hurst indices $H=0.05, 0.1,
\cdots, 0.3$, and $0.35$ are far less than 1. In these cases, the
recurrence networks are not  connected. So in the following
sections, we only consider connected networks with Hurst indices
$H$ from 0.4 to 0.95.

\section{Fundamental topological properties}

In this section, we numerically study the basic topological
features of recurrence networks including the degree distribution,
the clustering coefficient and the motif distribution. Recent
works have shown that a large number of real-world networks are
referred to as scale-free networks because the degree
distributions $P(k)$ follows a power-law
$$P(k) \sim k^{-\alpha}, \eqno{(4)} $$
where the degree exponent $\alpha$ varying in the range $2 <
\alpha < 3$ \cite{Barabasi99, Newman03, Barabasi03}. The degree
distributions of the visibility graphs constructed from FBMs have
power-law tails \cite{Lacasa08, Lacasa09, Ni09}. In addition, Xie
\textit{et al.} \cite{Xie11} found that horizontal visibility
graphs constructed from FBMs have exponential degree distributions
$$P(k) \sim e^{-\lambda k},   \eqno{(5)}$$
and the degree exponent $\lambda$ increases with Hurst index $H$.

It has been shown that many real-world complex networks share
similar universal statistical features, such as the small-world
character and the scale-free distribution, but may demonstrate
different local structure characteristics \cite{Watts98,
Barabasi99, Milo02}. An in-depth study of motif distribution can
help us to understand the design principles of complex networks on
the local structure level \cite{Milo02, Milo04}. In 2004, Milo
\textit{et al.} \cite{Milo04} developed an approach for comparing
network local structures, basing on the significance profile (SP).
They calculated the triad significance profile (TSP) for networks
from different fields and found that networks with similar
characteristic profiles are grouped into superfamilies.

\subsection{Degree distribution}

 Degree distribution is one of the most fundamental and
important topological properties of complex networks. We simulated
FBMs with different Hurst indices $H$ ranging from 0.40 to 0.95
with an increment of 0.05. For each value of $H$, we generated
1000 realizations and then calculated the average degree
distribution. For example, we give the average degree distribution
of recurrence networks constructed from FBMs with Hurst index
$H=0.6$ in Fig. 3. We find that all the recurrence networks
constructed from FBM series exhibit exponential degree
distributions. We estimated the degree exponents $\lambda$ and
display the average degree exponents $<\lambda>$ in Fig. 4. From
Fig. 4, we can see that the parameter $\lambda$ increases first
and then decreases with the increase of $H$. The inflection point
is at around $H=0.8$. We find that the relationship between $H$
and $<\lambda>$ can be fitted by a cubic polynomial function which
is also shown in Fig. 4.

\subsection{Clustering coefficient}

 The clustering coefficient measures the
density of triangles in a complex network. The definition of the
clustering coefficient given by Watts and Strogatz in Ref.
\cite{Watts98} has been used quite widely in complex network. The
local clustering coefficient $C_{i}$ is defined as
$$C_{i} = \frac{2E_{i}}{k_{i}(k_{i}-1)} = \frac{\sum_{j,h} A_{i,j}A_{j,h}A_{h,i}}{k_{i}(k_{i}-1)},   \eqno{(6)}$$
where $k_{i}$ is the degree of node $i$, $E_{i}$ is the actual
number of edges among
 $k_{i}$ nearest neighbor nodes of the node $i$. The clustering coefficient of the whole network is then given by the average
  of $C_{i}$ over all the nodes in the entire network:
$$C=\frac{1}{N}\sum_{i}C_{i}.   \eqno{(7)}$$
From the geometrical point of view, the local clustering
coefficient $C_{i}$ also can be defined as
$$C_{i} = \frac{number ~of ~triangles ~connected ~to ~vertex ~i}{number ~of ~triples ~centered ~on ~vertex ~i},   \eqno{(8)}$$
where a triple represents a vertex with edges connected any two
neighbor vertices. There are two different triples as shown in
Fig. 5. The triple $T_{1}$ will occur if the three vertices are
all close to each other. Conversely, the triple $T_{2}$ will occur
if the vertex $i$ is connected to its two neighbor vertices $j$
and $k$, but vertices $j$ and $k$ are not connected. In network
topology, the triangle structure $T_{1}$ means transitivity and
stability in the network.

In this paper, we study the dependence of the clustering
coefficient $C$ of the recurrence networks against the Hurst index
$H$ of the original FBMs. For each $H$, we simulated the FBM
$1000$ times, hence we got 1000 FBM series. For each FBM series, a
recurrence network was constructed and its clustering coefficient
$C$ was calculated. We display the relationship between $H$ and
the average clustering coefficient $<C>$ of recurrence networks in
Fig. 6. From Fig. 6, we can see that the average clustering
coefficient $<C>$ increases with the Hurst index $H$. The results
show that there are more triangle structures in the recurrence
network when the Hurst index $H$ increases from 0.4 to 0.95. This
means that the transitive and stable structure is more common for
large Hurst indices. This phenomenon can be easily explained by
following points.  It is well known that the Hurst index
characterizes the raggedness and irregularity of FBM. As already
mentioned, the FBM is a self-similar process with stationary
increments and possesses long-range dependence. The Hurst index
$H$ $(0<H<1)$ is a measure of the intensity of long-range
dependence in a FBM series. More specifically, the FBM with
$H=0.5$ indicates the absence of long-range dependence. And the
closer Hurst index $H$ is to $1$, the greater the extent of
long-range dependence or persistence. The Hurst index $H$ less
than $0.5$ corresponds to anti-persistence, which means that the
process displays strong negatively correlated and fluctuates
violently. For large Hurst index $H$, the FBM series is relatively
more smooth and regular. The result is that close data points in
the original FBM series are mapped to state vectors $X(i)$ which
are close to each other in a dense region of the phase space so
that these nodes $X(i)$ are more likely to be connected to each
other in recurrence network. As expected, there are many more
transitive and stable structures $T_{1}$ are to appear in
recurrence network constructed from the FBM series with large
Hurst index $H$. However, the FBM series with small Hurst index
$H$ is relatively more rough and irregular. As a consequence of
this, these close points in the original FBM series are mapped to
state vectors $X(i)$ in a sparse region of the phase space that
are less likely to be mutually connected, and therefore the triple
$T_{2}$ will be more common. From this point of view, the
recurrence network reflects the smoothness and regularity of the
FBMs to some extent.

\subsection{Motif distribution}

Because it is very time-consuming to study next a few properties
of large complex networks, we only generated 100 realizations of
FBM for each value of Hurst index $H$ and calculated the average
in the following.

Researches have shown that subgraphs or motifs are the building
blocks of complex networks and the superfamilies of networks can
be defined as the network motif patterns of occurrence
\cite{Milo02, Milo04}. Distinct from the degree distribution and
the clustering coefficient, the occurrence frequency of small
subgraphs or motifs can characterize the local structural
properties of real-world networks and then map the relation
between the function and the local structure of real-world systems
\cite{Milo02}. Xu \textit{et al.} \cite{Xu08} observed the
superfamily phenomenon in the nearest-neighbor networks
constructed from different time series. They also found that the
distribution of network motif ranks can be used to distinguish and
to characterize different types of dynamics in periodic, chaotic
and periodic with noise processes \cite{Xu08}. Xie \textit{et al.}
\cite{Xie11} calculated the frequencies of occurrence of the six
motifs of size $4$ within the horizontal visibility graphs
constructed from FBMs.

In this work, the constructed recurrence networks are  connected
and undirected, so here we only consider motifs with 4 nodes as in
Ref. \cite{Xie11}. Fig. 7 shows all six different network motifs
of size $4$ in connected and undirected networks. These motifs
reflects the different local structure characteristics in network.
There are two extreme blocks: motif $M_{1}$ and motif $M_{6}$. The
motif $M_{1}$ means the most irregular and nontransitive structure
that will appear if the node is connected to its three neighbor
nodes, but these neighbors are not connected to each other. On the
contrary, the motif $M_{6}$ indicates the most transitive
structure that will appear if the three neighbors are also
connected to each other. The relative occurrence frequencies of
motifs \cite{Xie11} are defined as
$$P(M) = \frac{n(M)}{\sum_{M=M_{1},...,M_{6}}n(M)},   \eqno{(9)}$$
where $n(M)$ is the number of motif $M$ in the network. After
constructing the recurrence networks based on the recurrence plot
method described above, we calculated the occurrence frequencies
of various motifs within the recurrence networks by the network
motif detection tool provided by Milo \textit{et al.} \cite{ht50}.
For each Hurst index $H$, which runs from 0.40 to 0.95 with a step
of 0.05, 100 realizations of FBMs were generated. Then we
constructed these time series into 100 recurrence networks and
obtained the average of $P(M)$.

In Fig. 8, we rearrange the relative occurrence frequencies
$<P(M)>$ in descending order for different Hurst indices $H$. As
we can see from Fig. 8, these seemingly unrelated recurrence
networks are divided into two superfamilies based on the very
similar motif rank distributions. Fig. 8(a) shows a superfamily
determined by the motif rank $M_{2}M_{3}M_{1}M_{5}M_{6}M_{4}$,
which includes the recurrence networks constructed from FBMs with
Hurst indices $H = 0.4, 0.45, 0.5, 0.55$, and $0.6$. The rest of
recurrence networks with Hurst indices $H = 0.65, \cdots, 0.95$
belong to another superfamily and the motif rank pattern is
$M_{2}M_{3}M_{5}M_{6}M_{1}M_{4}$ (Fig. 8(b)). In addition, three
particular motifs $M_{1}, M_{5}$, and $M_{6}$ play an important
role in the classification of recurrence networks we constructed
because their order determines the motif rank distributions. A
similar phenomenon was also observed in Refs. \cite{Xu08, Liu10}.
They gave a detailed explanation for this common phenomenon. Fig.
9 further illustrates the dependence of $<P(M)>$ with respect to
the Hurst index $H$. More specially, the occurrence frequency
$<P(M_{1})>$ of motif $M_{1}$ strictly monotonically decreases
with the increase of the Hurst index $H$ (see Fig. 9(a)). In
contrast to the motif $M_{1}$, the frequencies $<P(M_{5})>$ and
$<P(M_{6})>$ increase with the increase of the Hurst index $H$
(see Fig. 9(e) and Fig. 9(f)).
 We also note that the fluctuation of the
frequencies $<P(M_{4})>$ is comparatively small in Fig. 9(d),
which means that the motif $M_{4}$ almost does not appear in all
of the recurrence networks. These results indicate that there are
more regular and transitive structures in the networks when the
Hurst index $H$ varies from $0.4$ to $0.95$. That is to say, the
fully transitive motif $M_{6}$ will be common in these recurrence
networks constructed from the FBM series with large Hurst index
$H$ than the ones with small Hurst index $H$. On the contrary, the
irregular and nontransitive structure such as $M_{1}$ will be more
common for these network with small Hurst index $H$. These trends
are roughly coincide with the ones in the previous subsection. It
should be noted that two triples $T_{1}$ and $T_{2}$ in Fig. 5 are
just two motifs with $3$ nodes in undirected network. Measuring
and analyzing the density distribution of the motifs with $4$
nodes may be seen as a generalization of the aforementioned
transitivity concept. Therefore, as mentioned in the previous
subsection, we can give a similar explanation for these trends in
this subsection. In this sense, recurrence networks capture the
correlation of the associated FBMs.

\section{Fractal dimension}

In 1967, Mandelbrot introduced the fractal idea in Ref.
\cite{Mandelbrot67}. In fractal geometry, a fractal object is
self-similar because it contains small parts similar to the whole
\cite{Feder88, Mandelbrot83}. Recently, the fractal and
self-similarity properties of complex networks have been studied
extensively in various fields and systems \cite{Wang12, Palla05,
Goh06}. Song \textit{et al.} \cite{Song05} found that many
real-world networks such as the world-wide web (WWW), social
networks, protein-protein interaction (PPI) networks and cellular
networks consist of self-repeating patterns. They also believed
that these complex networks are self-similar under a certain
length-scale. After the small-world character and scale-free
property, self-similarity has become the third basic
characteristic of complex networks.

To gain further understanding of complex networks, numerous
algorithms have been developed to calculate the fractal dimension
of complex networks. Song \textit{et al.} \cite{Song07} proposed
to calculate the fractal dimension via a box-counting method. Kim
\textit{et al.} \cite{Kim07, Goh06} investigated the skeleton and
fractal scaling in complex networks using an improvement algorithm
which is a modified version of the original method introduced by
Song \textit{et al.} Then Zhou \textit{et al.} \cite{Zhou07}
developed an alternative algorithm, based on the edge-covering
box-counting, to detect self-similarity of cellular networks.
 In this section, we
adopt the random sequential box-covering method proposed by Kim
\textit{et al.} \cite{Kim07} for calculating the fractal
dimensions of recurrence networks.

For a given network, let $N_{B}(l_{B})$ be the smallest number of
boxes of lateral size $l_{B}$ which are needed to cover the entire
network. The fractal scaling implies the power-law relationship
between $N_{B}(l_{B})$ and $l_{B}$; the fractal dimension $d_{B}$
is then given by
$$N_{B}(l_{B}) \sim l_{B}^{-d_{B}}.   \eqno{(10)}$$
Usually, the fractal dimension $d_{B}$ can be obtained by fitting
the linear relationship between $N_{B}(l_{B})$ and $l_{B}$ in a
log-log plot.

Before introducing the random sequential box-covering algorithm
\cite{Kim07}, we use Floyd's algorithm \cite{Floyd62} of
Matlab-BGL toolbox \cite{Gleich} to calculate the shortest-path
distance matrix $D$ for each network according to the adjacency
matrix $A_{i,j}$ of the recurrence network. The random sequential
box-covering algorithm \cite{Kim07} can be described as follows.
We start with all vertices labelled as not burned. Then,
\begin{enumerate}

\item[(i)] Select a vertex randomly at each step; this vertex
serves as a seed and then consider the seed as the center of a
box.

\item[(ii)] For the center of the box, search all the neighbor
vertices within distance $l_{B}$ and burn all vertices
    which are found but have not been burned yet. Assign the newly burned vertices to the new box.
    If no newly burned vertex is found, then this box is discarded.

\item[(iii)] Repeat Steps (i) and (ii) until all vertices in the
entire network are assigned to their respective boxes.
\end{enumerate}

In this work, we simulated FBM series with different Hurst indices
$H$ ranging from 0.4 to 0.95 in the step of 0.05. For each $H$, we
simulated 100 FBM  time series and then constructed the obtained
series into 100 recurrence networks. Apparent power-law behaviors
of the two typical empirical recurrence networks constructed from
FBM series with different Hurst indices are shown in Fig. 10. The
fractal dimension $d_{B}$ is the absolute value of the slope of
linear regression between $\ln N_{B}(l_{B})$ and $\ln l_{B}$ for
each Hurst index. Fig. 11 shows the relationship between Hurst
index $H$ and the average fractal dimension $<d_{B}>$ over 100
realizations. As we can see from Fig. 11, the average fractal
dimension $<d_{B}>$ decreases with increasing $H$. Furthermore, it
is surprising that the curve shows a nice linear relationship:
$$<d_{B}> = 2.0064 - 1.0441H,   \eqno{(11)}$$
which approximates the theoretical relationship between Hurst
index $H$ and the fractal dimension $d$ of the graph of FBM $d = 2
- H$. Our numerical results show that the fractal dimension of the
recurrence networks constructed is very close to that of the graph
of the original FBMs. In other words, the fractality of FBMs is
inherited in their recurrence networks. We can explain here why
the two fractal dimensions are so closely related that the Hurst
index $H$ of the FBM series can be approximately estimated from
the perspective of the complex network.  As it has already been
mentioned, the pairs of nodes in the recurrence network can be
connected only when the two corresponding state vectors $X(i)$ in
phase space are close enough so that the phase space distance
between them is less than the critical recurrence threshold
$\varepsilon_{c}$. This is to say that recurrence networks we
constructed are random geometric graphs faithfully representing
the geometry of a set in phase space. Hence, neighborhood sizes
around a node in the network as well as around the corresponding
state vector in phase space will show the same scaling behavior.
This results in analogous behavior of the derived fractal
dimensions.

So far, we have numerically studied the basic topological
properties and  the fractal dimensions of the recurrence network.
More recently, some chaotic model systems such as logistic map,
H\'{e}non map, generalised baker's map and Bernoulli map, periodic
and two-dimensional quasiperiodic motions have been studied
analytically by Donner \textit{et al.} \cite{Donner11} and Donges
\textit{et al.} \cite{Donges12} from the point of view of
$\varepsilon$-recurrence network. To better understand the
interrelationships between network properties and dynamical
system, they defined some newly continuous measures based on
well-known graph theoretical measures and gave two novel notions
of dimension in phase space (clustering dimension and transitivity
dimension). They have put forward a theoretical framework for
these new measures. Their results indicated that
$\varepsilon$-recurrence networks show a strong relation between
dynamical systems and graph theory \cite{Donner11, Donges12}.
Their work provides us with a possible way to derive the
analytical results of these topological properties and dimensions
for recurrence network of FBM series in our future work.

\section{Multifractal analysis}

In real world, two fractal objects may have the same fractal
dimension but looks completely different. For these real-world
fractals, however, the tool of multifractal analysis shows better
performance and seems more powerful than fractal analysis.
Multifractal analysis has been successfully applied in a variety
of fields such as financial modelling \cite{Canessa00, Anh00},
biological systems (e.g. \cite{Yu01, Yu03, Yu04, Yu06}) and
geophysical data analysis (e.g. \cite{Yu09, Yu10}). Wang
\textit{et al.} \cite{Wang12} applied the modified fixed-size
box-counting algorithm to explore the multifractal behavior of
some theoretical networks, namely scale-free networks, small-world
networks, random networks, and a kind of real networks, namely PPI
networks of different species. Their numerical results indicate
that multifractality exists in scale-free networks and PPI
networks, while for small-world networks and random networks their
multifractality is not clear-cut.

Fixed-size box-covering algorithm \cite{Halsey86} is one of the
most common and important methods of multifractal analysis. For a
given measure $\mu$ with support set $E$ in a metric space, we
consider the partition sum
$$Z_{\epsilon}(q) = \sum_{\mu(B)\neq0}[\mu(B)]^{q},    \eqno{(12)}$$
where $q \in R$ and the sum runs over all different nonempty boxes
$B$ of a given size $\epsilon$ in a covering of the support set
$E$. From the definition above, we can  obtain $Z_{\epsilon}(q)
\geq 0$ and $Z_{\epsilon}(0) = 1$. The exponent $\tau_1(q)$ of the
measure $\mu$ can be defined as
$$\tau_1(q) = \lim_{\epsilon\rightarrow0}\frac{\ln Z_{\epsilon}(q)}{\ln \epsilon},   \eqno{(13)}$$
and the generalized fractal dimensions of the measure $\mu$ are
defined as
$$D(q) = \frac{\tau_1(q)}{q-1},   \eqno{(14)}$$
for $q \neq 1$, and
$$D(q) = \lim_{\epsilon\rightarrow0}\frac{Z_{1,\epsilon}}{\ln \epsilon}, \eqno{(15)}$$
for $q = 1$, where $Z_{1,\epsilon} = \sum_{\mu(B)\neq0}
\mu(B)\ln\mu(B)$. The linear regression of $[\ln
Z_{\epsilon}(q)]/(q-1)$ against $\ln \epsilon$ for $q \neq 1$
gives numerical estimates of the generalized fractal dimensions
$D(q)$, and similarly a linear regression of $Z_{1,\epsilon}$
against $\ln \epsilon$ for $q = 1$. In particular, $D(0)$ is the
box-counting dimension (or fractal dimension), $D(1)$ is the
information dimension, and $D(2)$ is the correlation dimension. If
the $\tau_1(q)$ or $D(q)$ curve versus $q$ is a straight line, the
object is monofractal. However, if this curve is convex, the
object is multifractal.

 In order to calculate the exponent $\tau_1(q)$ and the generalized fractal dimensions $D(q)$
 and then study the multifractality of networks, the measure $\mu$ of each box is usually defined as the ratio
 of the number of nodes covered by the box and the total number of nodes in the entire network. Wang \textit{et al.} \cite{Wang12}
 proposed a modified fixed-size box-counting algorithm to calculate the $\tau_1(q)$ and $D(q)$ and
 then investigate the multifractal behavior of complex networks. Recently Li {\it et al.} \cite{Libg2014} made some improvements
 based on the modified fixed-size box-counting algorithm proposed by Wang \textit{et al.} \cite{Wang12}.
 The improved algorithm \cite{Libg2014} can be summarized as follows:

\begin{enumerate}

\item[(i)] Initially, make sure all nodes in the entire network
are not covered and no node is  selected as a center of a box.

\item[(ii)] According to the size $N$ of networks constructed, set
$t = 1, 2, \cdots, 1000$ appropriately. Rearrange the nodes into
$1000$ different random orders. More specifically, in each random
order, nodes which will be selected as a center of a box are
randomly arrayed.

\item[(iii)] Set the radius $r$ of the box which will be used to
cover the nodes in the range $r \in [1, d]$, where $d$ is the
diameter of the network.

\item[(iv)] Treat the nodes in the $t$-th random order that we got
in (ii) as the center of a box, search all the neighbor nodes by
distance $r$ from the center and cover all nodes which are found
but have not been covered yet.

\item[(v)] If no newly covered nodes have been found, then this
box is discard.

\item[(vi)] Repeat Steps (iv) and (v) until all the nodes are
covered by the corresponding boxes. We denote the number of boxes
in this box covering as $N(t,r)$.

\item[(vii)] Repeat Steps (iv) to (vi) for all 1000 random orders
to find a box covering with minimal number of boxes $N(t,r)$.

\item[(viii)] For the nonempty boxes $B$ in the first box covering
with minimal number of boxes $N(t,r)$, define the measure of this
box as $\mu(B) = N_{B}/N$, where $N_{B}$ is the number of nodes
covered by the box $B$, and $N$ is the size of the network. then
calculate the partition sum $Z_{r}(q) = \sum_{\mu(B)\neq0}
[\mu(B)]^{q}$.

\item[(ix)] For different value of $r$, repeat Steps (iii) to
(viii) to calculate the partition sum $Z_{r}(q)$ and then use the
${Z_{r}(q)}$ for linear regression.
\end{enumerate}

A key step of linear regression is to obtain the appropriate range
of $r \in [r_{\min}, r_{\max}]$. Then we calculate the exponents
$\tau_1(q)$ and generalized fractal dimensions $D(q)$ in the
scaling ranges. In our calculation, we obtain the generalized
fractal dimensions through a linear regression of  $[\ln
{Z_{r}(q)}]/(q-1)$ against $\ln (r/d)$ for $q \neq 1$, and
similarly a linear regression of ${Z_{1,r}}$ against $\ln (r/d)$
for $q = 1$, where ${Z_{1,r}} = \sum_{\mu(B)\neq0} \mu(B) \ln
\mu(B)$.

Fig. 12 shows the linear regression for the recurrence network
constructed from FBM with Hurst index $H = 0.6$. The numerical
results show that the best fit occurs in the range $r \in (2,10)$
for this case ($H=0.6$). We select the best linear fit scaling
range to calculate the exponents $\tau_1(q)$ and generalized
fractal dimensions $D(q)$ and then to determine the
multifractality of recurrence networks from the shape of these
curves.

In this paper, we detect the multifractal behavior of recurrence
networks  using our improved version of the modified fixed-size
box-counting algorithm introduced by Li {\it et al.}
\cite{Libg2014}. For each value of the Hurst index $H$, we
averaged the results over $100$ realizations of the FBM series. We
summarize the corresponding numerical results in Table 1, which
includes the Hurst index $H$, the average box-counting dimension
$<D(0)>$, the average information dimension $<D(1)>$, the average
correlation dimension $<D(2)>$, and $\Delta D(q)$, where the
quantity
$$\Delta D(q) = \max D(q) - \min D(q)    \eqno{(16)}$$
was used to verify how $D(q)$ changes along each curve. From Table
1,  we can see that the average box-counting dimension $<D(0)>$,
the average information dimension $<D(1)>$ and the average
correlation dimension $<D(2)>$ are roughly decrease with the
increase of the Hurst index $H$ from $0.4$ to $0.95$. We show the
average $<\tau_1(q)>$ curves in Fig. 13 and average $<D(q)>$
curves in Fig. 14. From Figs. 13 and 14, we find that the
$<\tau_1(q)>$ and $<D(q)>$ curves of recurrence networks are not
straight lines. So the multifractality exists in these recurrence
networks constructed from FBM series. Meanwhile, we also find that
the multifractality of these networks becomes stronger at first
and then weaker which indicated by the value of $\Delta
D(q)=D(-10)-D(10)$ when the Hurst index of the associated time
series becomes larger from 0.4 to 0.95. In particular, the
recurrence network with the Hurst index $H=0.5$ possess the
strongest multifractality.  The dependence relationships of the
average information dimension $<D(1)>$ and the average correlation
dimension $<D(2)>$ on the Hurst index $H$ are given in Fig. 15. As
shown in Fig. 15, we find that these relationship can be well
fitted by following linear formulas:
$$<D(1)> = 1.7825 - 0.9046H,    \eqno{(18)}$$
and
$$<D(2)> = 1.5736 - 0.7956H.    \eqno{(19)}$$

\section{Conclusions}

In this work, we constructed recurrence networks from FBM series
based on the idea of recurrence plot. We extracted the statistical
properties of time series from the perspective of complex
networks. We studied the basic topological features of the
recurrence networks constructed from FBM series with different
Hurst indices $H$. Our numerical results indicate that the
recurrence networks constructed exhibit exponential degree
distributions. With the increase of $H$, the average degree
exponent $<\lambda>$ increases first and then decreases. The
relationship between $H$ and $<\lambda>$ can be fitted by a cubic
polynomial function. It was found that the average clustering
coefficient $<C>$ increases with the Hurst index $H$, which means
that there are more transitive and stable structures for networks
constructed from FBMs with large Hurst index $H$. At the
microscopic level, we investigated the motif rank distribution of
recurrence networks. We found that the recurrence networks are
grouped into two superfamilies based on the motif rank
distribution. The three key motifs $M_{1}, M_{5}$, and $M_{6}$
determine the motif rank pattern and then classify networks. At
the same time, we also paid attention to the dependence
relationship of the average occurrence frequency $<P(M)>$ with
respect to the Hurst index $H$.

From the aspect of fractality and self-similarity, we performed
fractal and multifractal analyses using the random sequential
box-covering method and the improved method based on modified
fixed-size box-counting algorithm, respectively. The numerical
results show that the average fractal dimension $<d_{B}>$
decreases with Hurst index $H$ and the linear relationship
$<d_{B}> \approx 2 - H$ was obtained surprisingly. Moreover, our
numerical results of multifractal analysis show that the
multifractality exists in these recurrence networks, and the
multifractality of these networks becomes stronger at first and
then weaker when the Hurst index of the associated time series
becomes larger from 0.4 to 0.95. In particular, the recurrence
network with the Hurst index $H=0.5$ possess  the strongest
multifractality. We also noted that the average information
dimension $<D(1)>$ and the average correlation dimension $<D(2)>$
are roughly decrease with the increase of the Hurst index $H$ from
$0.4$ to $0.95$. The dependence relationships of the average
information dimension $<D(1)>$ and the average correlation
dimension $<D(2)>$ on the Hurst index $H$ can be well fitted with
linear functions. From the above results, we conclude that the
inherent nature of time series affects the structure
characteristics of the associated networks and the dependence
relationship between them appears retained. Our works support that
complex networks is a suitable and effective tool to perform time
series analysis.

\section*{Acknowledgments}
This project was supported by the Natural Science Foundation of
China (Grant nos. 11071282 and 11371016), the Chinese Program for
Changjiang Scholars and Innovative Research Team in University
(PCSIRT) (Grant No. IRT1179), the Research Foundation of Education
Commission of Hunan Province of China (grant no. 11A122), the
Lotus Scholars Program of Hunan province of China, the Aid program
for Science and Technology Innovative Research Team in Higher
Educational Institutions of Hunan Province of China.

\newpage

\begin{table}
\caption{Comparison of recurrence networks transformed from FBM
series with different Hurst indexes. Here the average
  is calculated from 100 realizations.}
\begin{center}
\begin{tabular}{c|c|c|c|c}
\hline
$H$ & $<D(0)>$ &  $<D(1)>$ &  $<D(2)>$ & $\Delta D(q)$ \\
\hline
0.40 & 1.632955359455532 & 1.462113762291203 & 1.324568844472558 & 1.675426636215924    \\
0.45 & 1.595331264113667 & 1.345243849453671 & 1.182356286524482 & 2.067323904428986    \\
0.50 & 1.570768526526059 & 1.294972986568832 & 1.135291034040322 & 2.132106915333559    \\
0.55 & 1.496955506340691 & 1.276092858298725 & 1.120419709203580 & 1.778923958082282    \\
0.60 & 1.471252546702877 & 1.264224152311605 & 1.090156202522773 & 1.740584602742986    \\
0.65 & 1.348210637236297 & 1.225915988658060 & 1.103150737927043 & 1.642556252215174    \\
0.70 & 1.304446382642433 & 1.134347097612430 & 0.989670372719278 & 1.488149695525952    \\
0.75 & 1.251720498742988 & 1.085148989401195 & 0.950681208097028 & 1.445178739248894    \\
0.80 & 1.206437282010590 & 1.064415121034396 & 0.949594824553587 & 1.375606925727599    \\
0.85 & 1.154054167326104 & 1.014055134463063 & 0.904518905840645 & 1.231289283108065    \\
0.90 & 1.106243766095298 & 0.974358969483122 & 0.869233664007547 & 1.084781493689348    \\
0.95 & 1.056322443543536 & 0.921690208443068 & 0.819310269771388 & 1.042386684931838    \\
\hline
\end{tabular}
\end{center}
\end{table}

\begin{figure}
\centerline{\epsfxsize=10cm \epsfbox{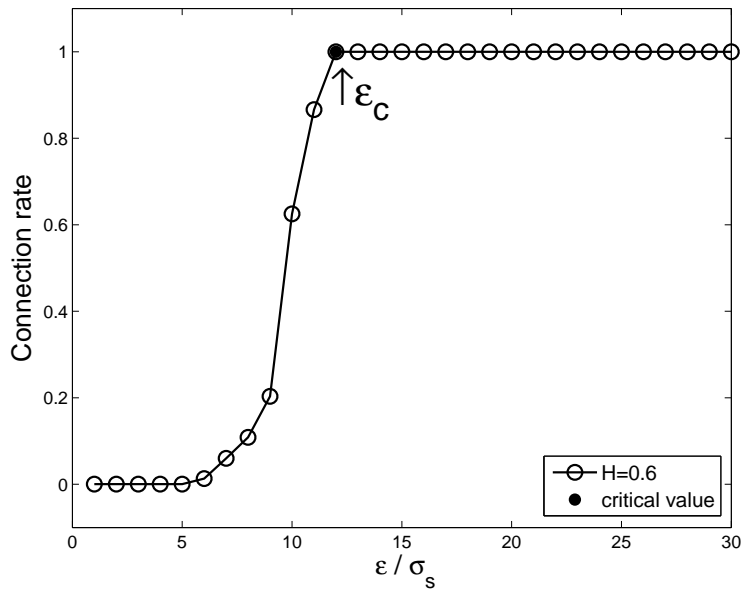}}
  \caption{The relationship between the recurrence threshold $\varepsilon$ and the connection rate of the recurrence network constructed from fractional Brownian motion series of length $L=2^{12}$ with Hurst index $H=0.6$.}
 \end{figure}

\begin{figure}

\centerline{\epsfxsize=10cm \epsfbox{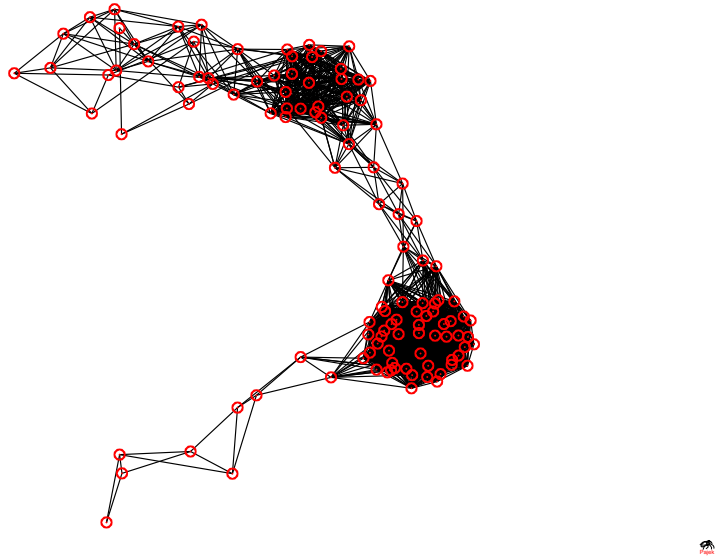}}
  \caption{ Recurrence network constructed from fractional Brownian motion
  series of length $L=2^8$ with Hurst index $H=0.6$.}
 \end{figure}

\begin{figure}
\centerline{\epsfxsize=10cm
\epsfbox{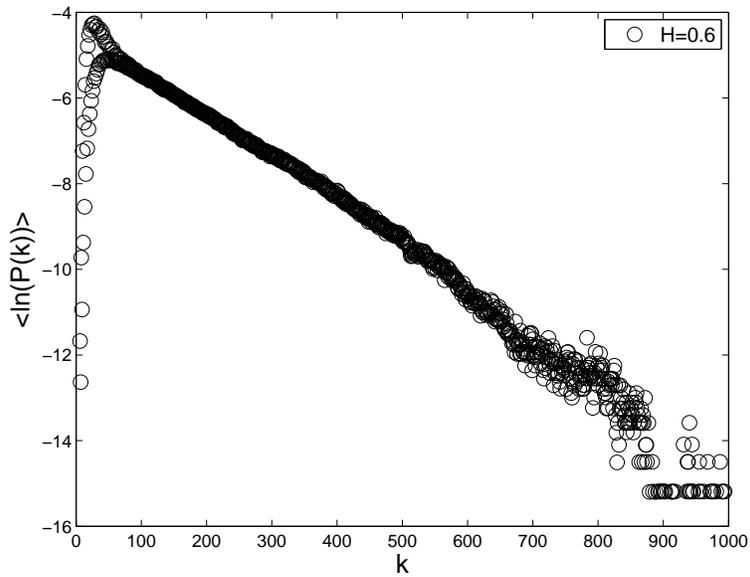}}
  \caption{Degree distribution of one recurrence network constructed from fractional Brownian motion with Hurst index $H=0.6$.}
 \end{figure}

\begin{figure}
\centerline{\epsfxsize=10cm
\epsfbox{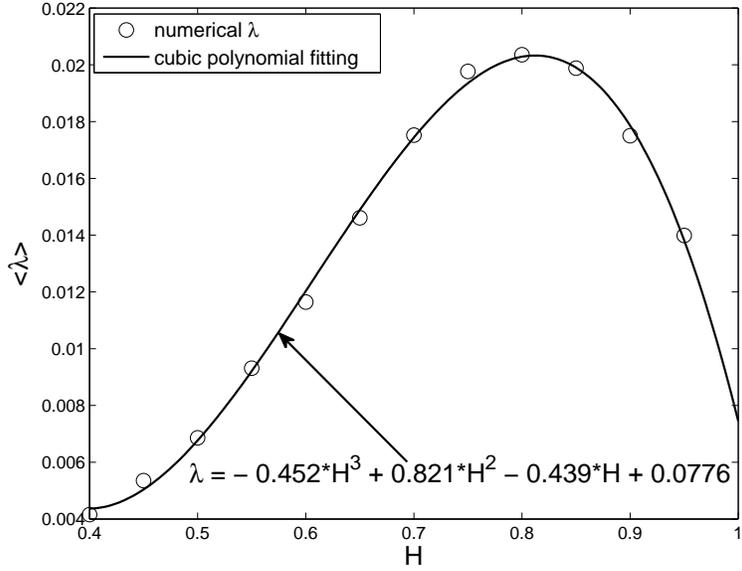}}
  \caption{The relationship between $H$ of fractional Brownian motion and average
  degree exponent $<\lambda>$ of the associated recurrence networks. Here the average is calculated from 1000 realizations.}
 \end{figure}

\begin{figure}
\centerline{\epsfxsize=10cm \epsfbox{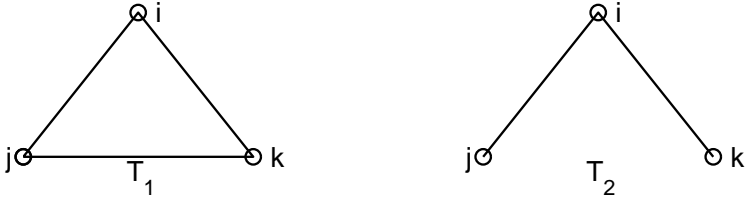}}
  \caption{All two triples centered on vertex $i$.}
 \end{figure}

\begin{figure}
\centerline{\epsfxsize=10cm \epsfbox{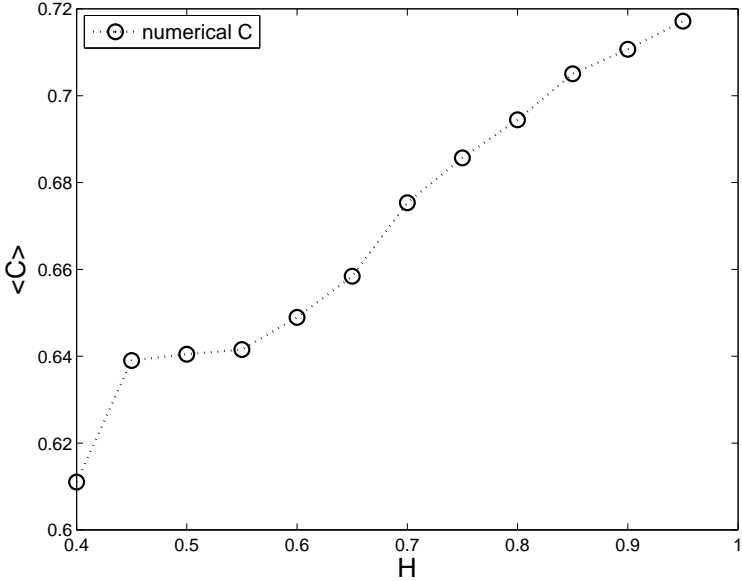}}
  \caption{The relationship between $H$ of fractional Brownian motion and average
  clustering coefficient $<C>$ of the associated recurrence networks. Here the average
  is calculated from 1000 realizations.}
 \end{figure}

\begin{figure}
\centerline{\epsfxsize=10cm \epsfbox{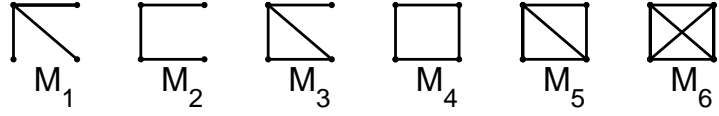}}
  \caption{All six network motifs of size $4$ in connected and undirected network. These motifs are
  labelled $M_{1}, M_{2}, M_{3}, M_{4}, M_{5}$, and $M_{6}$, respectively.}
 \end{figure}

\begin{figure}
\centerline{\epsfxsize=10cm
\epsfbox{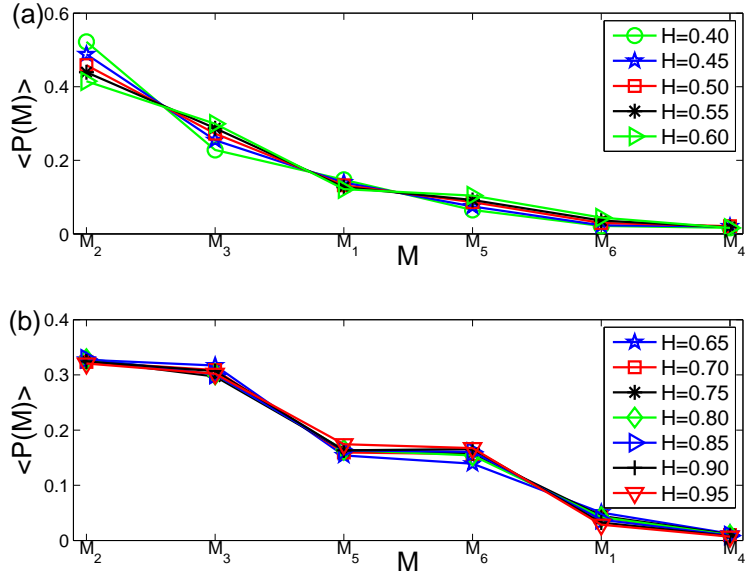}}
  \caption{ Network motif rank distributions of FBMs with different Hurst indices $H$.
  (a) Motif rank pattern $M_{2}M_{3}M_{1}M_{5}M_{6}M_{4}$ for $0.4 \leq H \leq 0.6$.
  (b) Motif rank pattern $M_{2}M_{3}M_{5}M_{6}M_{1}M_{4}$ for $0.65 \leq H \leq 0.95$.
  Here the average is calculated from 100 realizations.}
 \end{figure}

\begin{figure}
\centerline{\epsfxsize=10cm \epsfbox{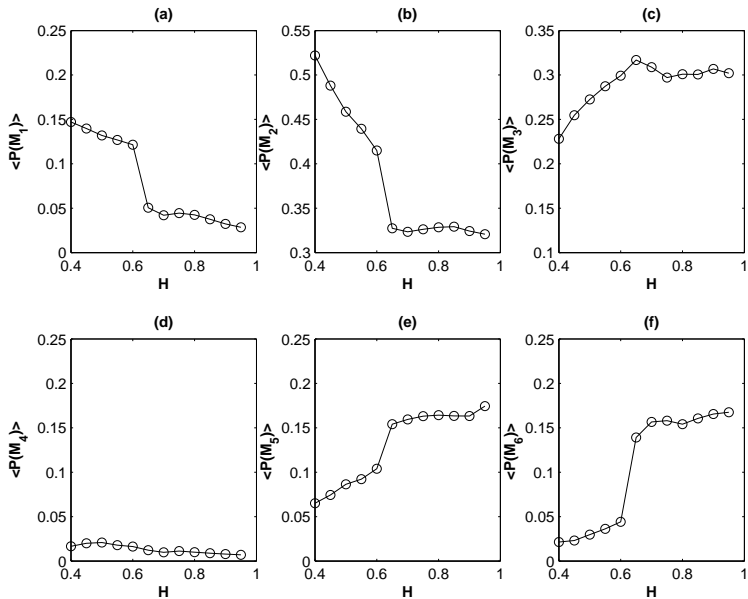}}
  \caption{Dependence of average occurrence frequencies of motifs $M_{1}, M_{2},
  M_{3}$, $M_{4}$, $M_{5}$, and $M_{6}$ on the Hurst indices $H$. Here the average
  is calculated from 100 realizations.}
\end{figure}

\begin{figure}
\centerline{\epsfxsize=10cm \epsfbox{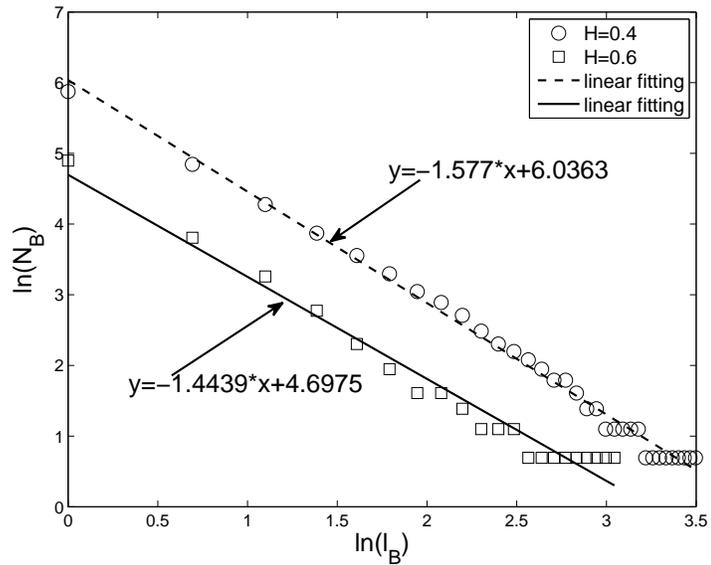}}
  \caption{The fractal scaling of recurrence networks constructed from FBMs with Hurst indices $H = 0.4$ and $0.6$.
  The fractal dimension is the absolute value of the slope of the linear fit.}
 \end{figure}

\begin{figure}
\centerline{\epsfxsize=10cm \epsfbox{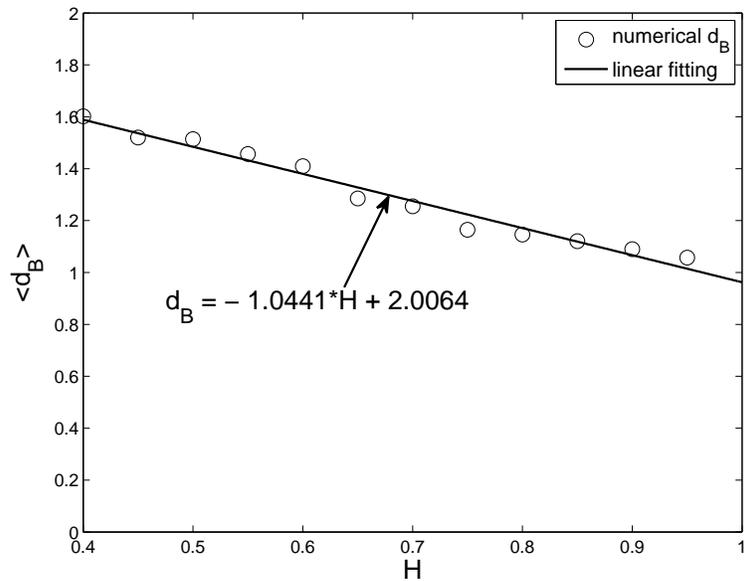}}
  \caption{The relationship between $H$ of fractional Brownian motion and average
  fractal dimension $<d_{B}>$ of the associated recurrence networks. Here the average
  is calculated from 100 realizations.}
 \end{figure}

\begin{figure}
\centerline{\epsfxsize=10cm
\epsfbox{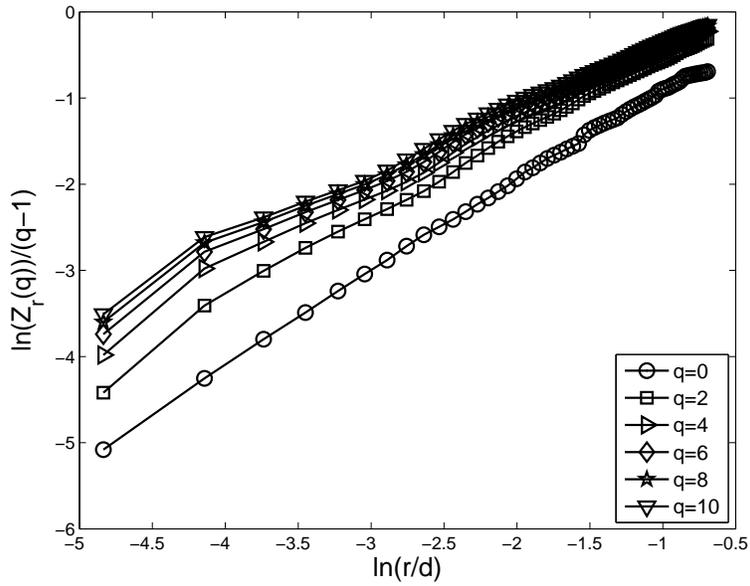}}
  \caption{Linear regressions for calculating the generalized dimensions of the recurrence network.}
 \end{figure}

\begin{figure}
\centerline{\epsfxsize=10cm \epsfbox{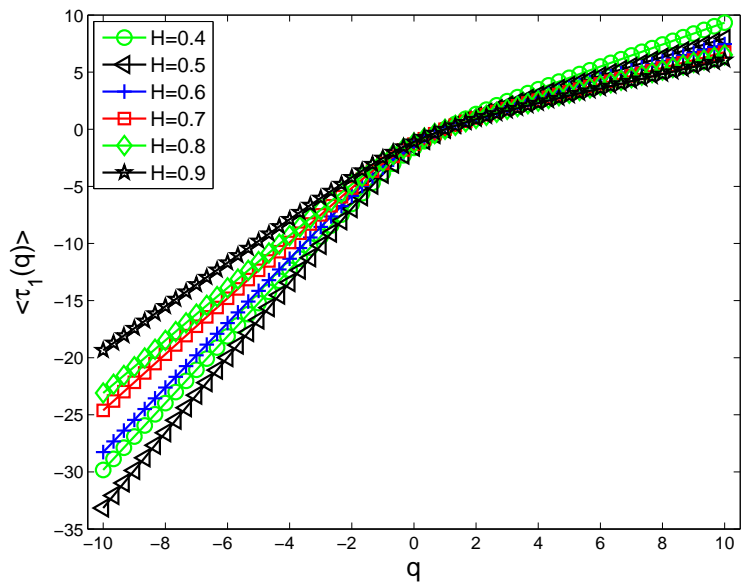}}
  \caption{ The average $\tau_1(q)$ curves of the recurrence networks. Here the average
  is calculated from 100 realizations.}
 \end{figure}

\begin{figure}
\centerline{\epsfxsize=10cm \epsfbox{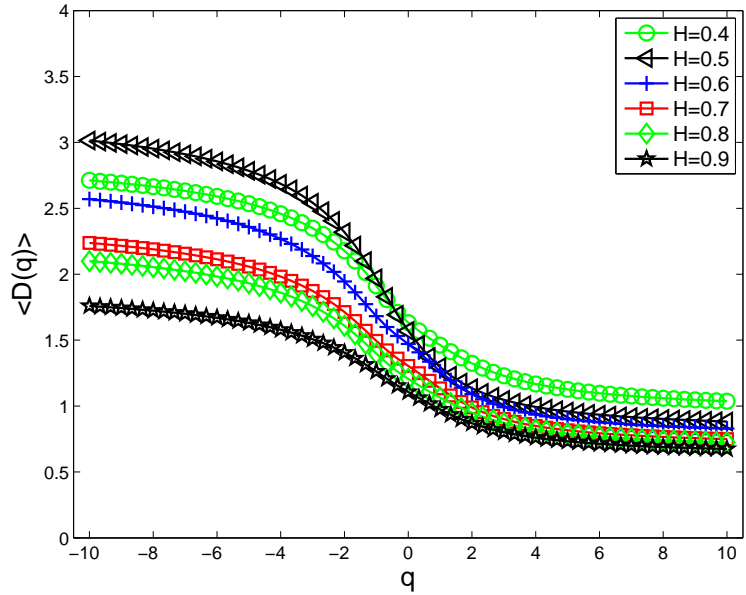}}
  \caption{ The average $D(q)$ curves of the recurrence networks, $D(0)$ is the fractal dimension of the network. Here the average
  is calculated from 100 realizations.}
 \end{figure}

\begin{figure}
\centerline{\epsfxsize=10cm \epsfbox{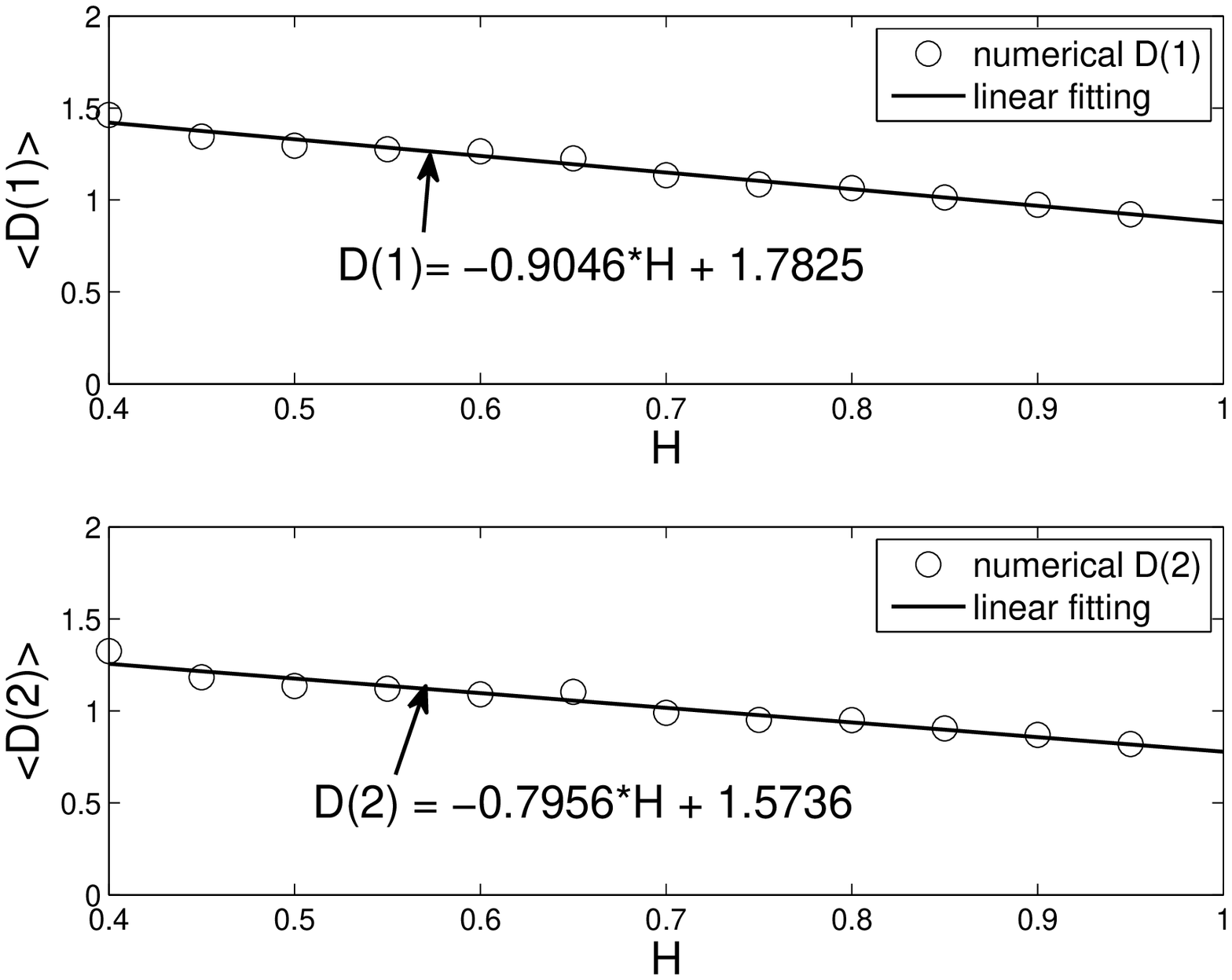}}
  \caption{The dependence of the average information dimension $<D(1)>$ and the average correlation dimension $<D(2)>$ of the associated recurrence networks with respect to the Hurst index $H$ of fractional Brownian motion. Here the average
  is calculated from 100 realizations.}
 \end{figure}


\begin{thebibliography}{10}

{\footnotesize

\bibitem[1] {Albert02} R. Albert and A.L. Barab\'{a}si, \emph{Rev. Mod. Phys.}, \textbf{74} (2002) 47-97.

\bibitem[2] {Donner12} R.V. Donner, Y. Zou, J.F. Donges, N. Marwan, and J. Kurths, \emph{Phys. Rev. E}, \textbf{81} (2010) 015101(R).

\bibitem[3] {Donner10} R.V. Donner, Y. Zou, J.F. Donges, N. Marwan, and J. Kurths, \emph{New J. Phys.}, \textbf {12} (2010) 033025.

\bibitem[4] {Watts98} D.J. Watts and S.H. Strogatz, \emph{Nature}, \textbf{393(6684)} (1998) 440-442.

\bibitem[5] {Barabasi99} A.L. Barab\'{a}si and R. Albert, \emph{Science}, \textbf{286(5439)} (1999) 509-512.

\bibitem[6] {Newman03} M.E.J. Newman, \emph{SIAM Rev}, \textbf{45} (2003) 167-256.

\bibitem[7] {Xu09} M. Small, J. Zhang, and X.K. Xu, \emph{in: Lect. Notes Institue
Comput. Sci. Soc. Infor. Telecommun. Engineering}, \textbf{5}
(2009) 2078-2089.

\bibitem[8] {Lacasa08} L. Lacasa, B. Luque, F. Ballesteros, J. Luque, and J.C. Nu\~{n}o, \emph{Proc. Natl. Acad. Sci. USA.}, \textbf {105} (2008) 4972-4975.

\bibitem[9] {Luque09} B. Luque, L. Lacasa, F. Ballesteros, and J. Luque, \emph{Phys. Rev. E}, \textbf{80} (2009) 046103.

\bibitem[10] {Li08} C.B. Li, H. Yang, and T. Komatsuzaki, \emph{Proc. Natl. Acad. Sci. USA.}, \textbf{105} (2008) 536-541.

\bibitem[11] {Marwan09} N. Marwan, J.F. Donges, Y. Zou, R.V. Donner, and J. Kurths, \emph{Phys. Lett. A}, \textbf{373} (2009) 4246-4254.

\bibitem[12] {Xu08} X.K. Xu, J. Zhang, and M. Small, \emph{Proc. Natl. Acad. Sci. USA.}, \textbf{105} (2008) 19601-19605.

\bibitem[13] {Liu10} C. Liu and W.X. Zhou, \emph{J. Phys. A}, \textbf{43} (2010) 495005.

\bibitem[14] {Eckmann87} J.P. Eckmann, S.O. Kamphorst, and D. Ruelle, \emph{Europhys. Lett.}, \textbf{5} (1987) 973-977.

\bibitem[15] {Zbilut92} J.P. Zbilut and C.L. Webber Jr., \emph{Phys. Lett. A}, \textbf{171(3-4)} (1992) 199-203.

\bibitem[16] {Webber94} C.L. Webber Jr. and J.P. Zbilut, \emph{J. Appl. Physiol.}, \textbf{76(2)} (1994) 965-973.

\bibitem[17] {ht21} TOCSY: http://tocsy.pik-potsdam.de

\bibitem[18] {Zbilut98} J.P. Zbilut, A. Giuliani, and C.L. Webber Jr., \emph{Phys. Lett. A}, \textbf{246(1-2)} (1998) 122-128.

\bibitem[19] {Marwan02} N. Marwan and J. Kurths, \emph{Phys. Lett. A}, \textbf{302(5-6)} (2002) 299-307.

\bibitem[20] {Romano04} M.C. Romano, M. Thiel, J. Kurths, and W.von Bloh, \emph{Phys. Lett. A}, \textbf{330(3-4)} (2004) 214-223.

\bibitem[21] {Donges12} J.F. Donges, J.Heitzig, R.V. Donner, and J. Kurths, \emph{Phys. Rev. E}, \textbf{85} (2012) 046105.

\bibitem[22] {Donner10a} R.V. Donner, J.F. Donges, Y. Zou, N. Marwan, and J. Kurths, In: \emph{Proceedings of the 2010 International
Symposium on Nonlinear Theory and its Applications NOLTA2010
(Krakow 2010)}, Tokyo : IEICE 2010, p87-90.

\bibitem[23] {Mandelbrot68} B.B. Mandelbrot and J.W. Van Ness, \emph{SIAM Rev.}, \textbf{10(4)} (1968) 422-437.

\bibitem[24] {Lacasa09} L. Lacasa, B. Luque, J. Luque, and J.C. Nu\~{n}o, \emph{Europhys. Lett.}, \textbf{86} (2009) 30001.

\bibitem[25] {Xie11} W.J. Xie and W.X. Zhou, \emph{Physica A}, \textbf{390} (2011) 3592-3601.

\bibitem[26] {ZLY2013} Y. Zhou, Y. Leung, and Z.G. Yu, \emph{Phys. Rev. E}, \textbf{87} (2013) 012921.

\bibitem[27] {Riley1999} M.A. Riley, R. Balasubramaniam and M.T. Turvey, \emph{Gait and Posture}, \textbf{9} (1999) 65-78.

\bibitem[28] {Kim07} J.S. Kim, K.I. Goh, G. Salvi, E. Oh, B. Kahng, and D. Kim, \emph{Phys. Rev. E}, \textbf{75} (2007) 016110.

\bibitem[29] {Libg2014}  B.G. Li, Z.G. Yu  and Y. Zhou, \emph{J. Stat. Mech.: Theor. Exp.}, \textbf{2014} (2014) P02020.

\bibitem[30] {Wang12} D.L. Wang, Z.G. Yu, and V. Anh, \emph{Chin. Phys. B}, \textbf{21(8)} (2012) 080504.

\bibitem[31] {Packard80} N.H. Packard, J.P. Crutchfield, J.D. Farmer, and R.S. Shaw, \emph{Phys. Rev. Lett.}, \textbf{45(9)} (1980) 712-716.

\bibitem[32] {Takens81} F. Takens, \emph{in: Rand D.A.,
Young L.S.(Eds), Dynamical Systems and Turbulence, Lecture Notes
in Mathematics, Berlin: Springer-Verlag}, \textbf{898} (1981)
366-381.

\bibitem[33] {Casdagli91} M. Casdagli, S. Eubank, J.D. Farmer, and J. Gibson, \emph{Physica D}, \textbf{51} (1991) 52-98.

\bibitem[34] {Rosenstein94} M.T. Rosenstein, J.J. Collins, and C.J. De Luca, \emph{Physica D}, \textbf{73(1-2)} (1994) 82-98.

\bibitem[35] {Fraser86} A.M. Fraser and H.L. Swinney, \emph{Phys. Rev. A}, \textbf{33(2)} (1986) 1134-1140.

\bibitem[36] {Kennel92} M.B. Kennel, R. Brown, and H.D. Abarbanel, \emph{Phys. Rev. A}, \textbf{45(6)} (1992) 3403-4311.

\bibitem[37] {Cao97} L.Y. Cao, \emph{Physica D}, \textbf{110(1-2)} (1997) 43-50.

\bibitem[38] {Marwan07} N. Marwan, M.C. Romano, M. Thiel, and J. Kurths, \emph{Phys. Rep.}, \textbf{438(5-6)} (2007) 237-329.

\bibitem[39] {Riley05} M.A. Riley and G.C. Van Orden, \emph{Tutorials in contemporary nonlinear
methods for the behavioral sciences}
(http://www.nsf.gov/sbe/bcs/pac/nmbs/nmbs.jsp) (2005).

\bibitem[40] {Erdos59} P. Erd\H{o}s and A. R\'{e}nyi, \emph{Publicationes Mathematicae Debrencen}, \textbf{6} (1959) 290.

\bibitem[41] {Roulston99} M.S. Roulston, \emph{Physica D}, \textbf{125(3-4)} (1999) 285-294.

\bibitem[42] {Tang12} Z.H. Tang and Z.G. Yu, \emph{Chin. J. Eng. Math.}, \textbf{29(4)} (2012) 499-506.

\bibitem[43] {Gleich} Gleich D.F., \emph{A graph library for Matlab based on the boost graph library},
http://dgleich.github.com/matlab-bgl

\bibitem[44] {ht40} http://pajek.imfm.si/doku.php?id=pajek

\bibitem[45] {Barabasi03} A.L. Barab\'{a}si and E. Bonabeau, \emph{Scientific American}, (2003) 50-59.

\bibitem[46] {Ni09} X.H. Ni, Z.Q. Jiang, and W.X. Zhou, \emph{Phys. Lett. A}, \textbf{373} (2009) 3822-3826.

\bibitem[47] {Milo02} R. Milo, S.S. Orr, S. Itzkovitz, N. Kashtan, D. Chklovskii, and U. Alon, \emph{Science}, \textbf{298} (2002) 824-827.

\bibitem[48] {Milo04} R. Milo, S. Itzkovitz, N. Kashtan, R. Levitt, S.S. Orr, I. Ayzenshtat, M. Sheffer, and U. Alon, \emph{Science}, \textbf{303} (2004) 1538-1542.

\bibitem[49] {ht50} http://www.weizmann.ac.il/mcb/UriAlon/

\bibitem[50] {Mandelbrot67} B.B. Mandelbrot, \emph{Science}, \textbf{155} (1967) 636-638.

\bibitem[51] {Feder88} J. Feder, \emph{Fractals} (Plenum, New York, 1988).

\bibitem[52] {Mandelbrot83} B.B. Mandelbrot, \emph{The Fractal Geometry of Nature} (New York: Academic Press 1983).

\bibitem[53] {Palla05} G. Palla, I. Derenyi, I. Farkas, and T. Vicsek, \emph{Nature}, \textbf{435(7043)} (2005) 814-818.

\bibitem[54] {Goh06} K.I. Goh, G. Salvi, B. Kahng, and D. Kim, \emph{Phys. Rev. Lett.}, \textbf{96(1)} (2006) 018701.

\bibitem[55] {Song05} C. Song, S. Havlin, and H.A. Makse, \emph{Nature}, \textbf{433} (2005) 392-395.

\bibitem[56] {Song07} C. Song, L.K. Gallos, S. Havlin, and H.A. Makse, \emph{J. Stat. Mech.: Theor. Exp.}, \textbf{3} (2007) P03006.

\bibitem[57] {Zhou07} W.X. Zhou, Z.Q. Jiang, and D. Sornette, \emph{Physica A}, \textbf{375(2)} (2007) 741-752.

\bibitem[58] {Floyd62} R.W. Floyd, \emph{Commun. ACM}, \textbf{5(6)} (1962) 345.

\bibitem[59] {Donner11} R.V. Donner, J. Heitzig, J.F. Donges, Y. Zou, N. Marwan, and J. Kurths, \emph{Eur. Phys. J. B}, \textbf{84} (2011) 653-672.

\bibitem[60] {Canessa00} E. Canessa, \emph{J. Phys. A: Math. Gen.}, \textbf{33} (2000) 3637-3651.

\bibitem[61] {Anh00} V.V. Anh, Q.M. Tieng, and Y.K. Tse, \emph{Int. Trans. Oper. Res.}, \textbf{7} (2000) 349-363.

\bibitem[62] {Yu01} Z.G. Yu, V. Anh, and K.S. Lau, \emph{Phys. Rev. E}, \textbf{64} (2001) 031903.

\bibitem[63] {Yu03} Z.G. Yu, V. Anh, and K.S. Lau, \emph{Phys. Rev. E}, \textbf{68} (2003) 021913.

\bibitem[64] {Yu04} Z.G. Yu, V. Anh, and K.S. Lau, \emph{J. Theor. Biol.}, \textbf{226} (2004) 341-348.

\bibitem[65] {Yu06} Z.G. Yu, V. Anh, and K.S. Lau, \emph{Phys. Rev. E}, \textbf{73} (2006) 031920.

\bibitem[66] {Yu09} Z.G. Yu, V. Anh, and R. Eastes, \emph{J. Geophys. Res.}, \textbf{114} (2009) A05214.

\bibitem[67] {Yu10} Z.G. Yu, V. Anh, Y. Wang, D. Mao, and J. Wanliss, \emph{J. Geophys. Res.}, \textbf{115} (2010) A10219.

\bibitem[68] {Halsey86} T.C. Halsey, M.H. Jensen, L.P. Kadanoff, I. Procaccia, and B.I. Shraiman, \emph{Phys. Rev. A}, \textbf{33} (1986) 1141-1151.

}

\end{thebibliography}
\end{document}